\documentclass{edp-conf}
\usepackage{graphicx}
\begin{document}
\def\la{\buildrel<\over\sim}
\def\ga{\buildrel>\over\sim}

\TitreGlobal{SF2A 2005}

\title{ON THE FREQUENCY OF FIELD GALACTIC Be STARS}

\author{Zorec, J.}\address{Institut d'Astrophysique de Paris, UMR7095 CNRS, 
Universit\'e Pierre \& Marie Curie}
\author{Fr\'emat, Y.}\address{Royal Observatory of Belgium}

\runningtitle{Frequency of Be stars}

\index{Zorec, J.}
\index{Fr\'emat, Y.}

\maketitle

\begin{abstract}Since Be stars belong to the high velocity tail of a single B
star velocity distribution, the appearance of the Be phenomenon must be 
independent of the stellar mass. In the present paper we show that the shape
of the distribution of the number fraction N(Be)/N(Be+B) against the spectral
type can be explained in terms of the Balmer line emission efficiency as a 
function of the effective temperature. 
\end{abstract}

\section{Introduction}

 Field Be stars near the Sun represent a sample of objects in different stages 
of the main sequence evolutionary phase (MS). Several hypothesis have been put 
forward to explain the distribution of the frequency of Be stars against the 
spectral type. As Be stars belong to the high velocity tail of a single B star 
velocity distribution (Zorec et al. 2005, Martayan et. al. 2005), there is no
physical reason that the fraction of Be stars be mass-dependent as apparently
it comes up from statistics. 

\section{Method}

 In this contribution we show that the observed distribution of the frequency 
of Be stars is determined by the Balmer line emission efficiency as a function 
of the effective temperature. The probability of detecting a Be star can be 
written as:

\begin{equation}
{\rm d}N(Be,i) \propto P_{\rm Be}E_{H\alpha}(T_{\rm eff},\tau,i)\phi(\tau
)N(T_{\rm eff})\sin i{\rm d}\tau{\rm d}i
\label{prob1} 
\end{equation}
 
\noindent where $P_{\rm Be}$ is the probability that a B star becomes Be; 
$E_{H\alpha}$ is the average intensity of the H$\alpha$ emission produced by 
a circumstellar disc of opacity $\tau$ in the center of the H$\alpha$ line 
seen under an aspect angle $i$; $\phi(\tau)$ is the probability the disc had
an opacity $\tau$; $N(T_{\rm eff)}$ is the total number of stars having 
effective temperature $T_{\rm eff}$; $\sin i$ is the probability of seeing the 
disc at an inclination $i$. The emission intensity factor is $ E_{H\alpha
}(T_{\rm eff},\tau,i)=$ $S_{H\alpha}\Delta(i)(1-e^{-{\tau\over\cos i}})$, 
where $S_{H\alpha}(T_{\rm eff})$ is the photoionization-, thus, $T_{\rm 
eff}$-dominated H$\alpha$ line source function; $\Delta(i)$ contains 
geometrical parameters of the disc, which we assume be the same for all stars. 
The expected number of Be $N(Be)$ is then given by $N(Be)=$ $\beta F(T_{\rm 
eff},\tau_o)N(B+Be)$, where $N(B+Be)$ is the number of all stars in the 
studied mass or $T_{\rm eff}$ interval. Since we assume $P_{\rm Be}(M)=$ 
$constant$, it comes up that $\beta=$ $constant$ which implies the number of
expected Be stars in a given mass range is not mass-dependent.  

\section{Results and Conclusions}

 Figure 1a shows $\overline{E(T_{\rm eff},\tau)}$ against $T_{\rm eff}$ and
$\tau_o$. To represent $N(M)$ we used the total counts of B and Be stars per
mass interval in the V = 7 magnitude limited volume around the Sun, so as 
to reflect closely the local IMF. Figure 1b shows the histogram of the 
observed fractions of Be stars and the histogram predicted by (2.1) with 
$\tau_o = 1$. Distributions resemble each other closely and maxima at B2 and 
from B7/8 are well accounted for. The maximum at B2 is due to the high 
H$\alpha$ emission, while that at B7/8 is due to the increase of the IMF 
function. The slight excess of predicted Be stars in the cool extreme of the 
distribution can be due to stars that have not yet developed the Be phenomenon 
and to actual Be stars with too tiny or cool discs whose emission has remained 
unseen. There is enough fast rotating B stars, called Bn stars, which can fit 
the excess predicted in the low temperature side (Zorec 2004).

\begin{figure}[h]
\centering
\includegraphics[height=5cm,width=11cm]{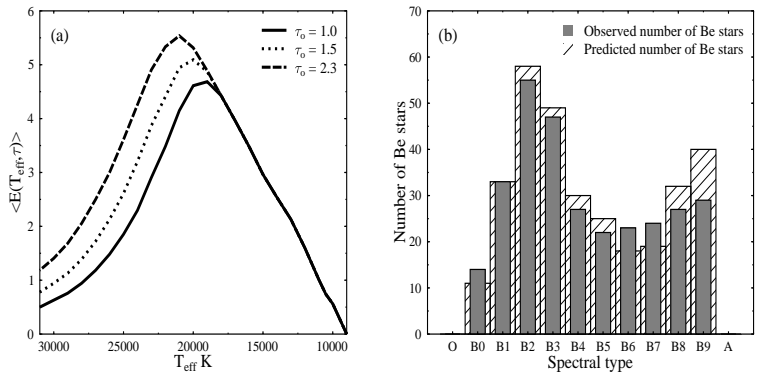}
\caption{(a): Emission intensity function in the H$\alpha$ line as a function
of $T_{\rm eff}$ and $\tau_o$. (b) Observed and predicted number of Be stars}
\end{figure}


\begin{thebibliography}{}
\bibitem{}Martayan, C., Zorec, J., Fr\'emat, Y. et al. 2005, SF2A/2005, this 
issue
\bibitem{}Zorec J., Martayan, C., Fr\'emat, Y. et al. 2005, in: Active 
OB-Stars: Laboratories for Stellar and Circumstellar Physics, Workshop
\bibitem{}Zorec J. 2004, in: Stellar Rotation, IAU Symp. No 215, p. 73
\end{thebibliography}
\end{document}